\documentclass[titlepage,aps,preprint,floatfix]{revtex4-2}
\usepackage{graphicx}
\usepackage{fancyhdr}
\usepackage{amsmath,amsfonts,color,shadow}
\usepackage{upgreek}
\usepackage{tabularx}
\begin{document}
\title{On Analysis of Laser Plasma Aluminum Monoxide Emission Spectra}
\author{Christian G. Parigger}
\address{Physics and Astronomy Department, University of Tennessee,  University of Tennessee Space Institute,
Center for Laser Applications, \\ 411 B.H. Goethert Parkway, Tullahoma, TN 37388-9700, USA; cparigge@tennessee.edu}

\begin{abstract}
\noindent This work communicates analysis of aluminum monoxide, AlO, laser-plasma emission records using line strength data and the ExoMol astrophysical database. A nonlinear fitting program computes comparisons of measured and simulated diatomic molecular spectra. Predicted cyanide spectra of the AlO, ${\rm B}\  ^2\,\Sigma^+ \longrightarrow {\rm X} \ ^2\,\Sigma^+$, $\Delta {\rm v} = 0, \pm 1, \pm 2, + 3$ sequences and progressions compare nicely with 1 nanometer resolution experimental results. The analysis discusses experiment data captured during laser ablation of Al$_2$O$_3$ with 266-nm, 6-mJ pulses. The accuracy of the AlO line strength data is better than one picometer. This work presents as well comparison of the $^{27}$Al$^{16}$O line strength and of ExoMol data for spectral resolutions of 0.1\,nm and 0.07\,nm. Accurate AlO databases show a volley of applications in laboratory and astrophysical plasma diagnosis.


\end{abstract}

\maketitle


\section{Introduction}

The diatomic molecule aluminum monoxide, AlO, occurs in plasma-emission following laser ablation of aluminum containing samples, including alumina (Al$_2$O$_3$) \cite{Dors} or aluminum containing alloys \cite{Surmick}. Combustion of aluminized propellants also reveals nice AlO flame emission spectra \cite{Sandia}. Usually, accurate diatomic line strengths data are preferred in the analysis \cite{SAA,Foundations1,Pariggerbook} of recorded data. However, recent interest in exo-planet spectroscopy motivates determination of molecular databases, viz. ExoMol \cite{ExoMol}. The ExoMol database lists various AlO isotopologues, however this work focuses on $^{27}$Al$^{16}$O. The transition of interest is the
AlO $B \, ^2\Sigma^+ - X \, ^2\Sigma^+$ band system that is similar in principle to previously communicated cyanide, CN $B \, ^2\Sigma^+ - X \, ^2\Sigma^+$ band system \cite{CNSubmitted}.

Spectroscopy \cite{Kunze,Fujimoto,Ochkin,Demtr1,Demtr2,Miziolek,SinghThakur} of laser-plasma reveals clean AlO band system for delays of the order of several dozens of microseconds from the initial ablation plasma generation using pulse widths of a few nanoseconds.
For aluminum monoxide spectroscopy, one can employ the ExoMol database in conjunction with the PGOPHER program for simulating rotational, vibrational and electronic spectra \cite{PGOPHER}. There are of course other databases that can be accessed \cite{McKemmish} for diatomic molecules, including HITEMP that for example shows hydroxyl, OH, data \cite{HITEMP}. The ExoMol AlO data files for the states and the transitions are converted in this work to line strength files for the purpose of utilizing previously communicated and extensively tested line-strength data that are freely available along with MATLAB \cite{MATLAB} scripts for a subset of transitions associated with the AlO B-X band systems \cite{SAA,Foundations1,Pariggerbook}.



\section{Experimental and Analysis Overview}

The data from laser ablation experiments with frequency quadrupled Q-switched Nd:YAG radiation \cite{Dors} show a range of 430 nm to 540 nm, and the published comparisons with line strength data reveal a temperature of 3,432 Kelvin at a delay of 20~$\upmu$s. The measurements use standard laser-induced-breakdown-spectroscopy (LIBS) equipment. The analysis in that work utilizes AlO-lsf line strengths and the Nelder-Mead downhill simplex, non-linear fitting algorithm \cite{NMalgorithm}. The analysis communicated in this work is designed such that the same nonlinear method can be used, but the ExoMol data base for AlO is recast in a set of transitions with line strength data that are determined from Einstein A-coefficients.

\subsection{Diatomic Molecular Analysis}

The computation of diatomic molecular spectra utilizes line strength data. The Boltzmann equilibrium spectral program (BESP) and the Nelder-Mead temperature (NMT) program allow one to respectively compute an emission spectrum and fit theoretical to experimental spectra. The construction of the communicated molecular AlO line strengths ``AlO-lsf'' \cite{Foundations1}  first, makes use of Wigner-Witmer eigenfunctions and a diatomic line position fitting program, second, computes Frank-Condon factors and r-centroids, and third, combines these factors with the rotational factors that usually decouple from the overall molecular line-strength due to the symmetry of diatomic molecules.  In turn, the ExoMol states and transition files for AlO \cite{LineListAlO,ExoMolAlO} are utilized for the generation of line strength data that can be used with BESP and NMT.

The ExoMol data show Einstein A-coefficients that are converted to line strengths \cite{Condon,Hilborn,Thorne}, S, for electric dipole transitions, using

\begin{equation}
A_{ul}= \frac{16 \pi^3}{3 g_u h \upepsilon_0 \uplambda^3} \left(e\ a_0\right)^2 S_{ul}, \ \ \ \ \ \ \ g_u = 2 ( 2 J_u +1).
\label{equationA}
\end{equation}

\noindent Here, $A_{ul}$ denotes the Einstein A-coefficient for a transition from an upper, $u$, to a lower, $l$, state, and $h$ and $\varepsilon_0$ are Planck's constant and vacuum permittivity, respectively. The elementary charge is $e$, the Bohr radius is $a_0$, and  the transition strength is $S_{ul}$. The line strength, $S$, that is used in the MATLAB scripts is expressed in traditional spectroscopy units (stC$^2$ cm$^2$). The wavelength of the transition is $\uplambda$, $g_u$ is the upper state degeneracy and $J_u$ the total angular momentum of the upper state. In the establishment of line strength data, Hund's case (a) basis functions are preferred in connection with application of the Wigner and Witmer \cite{Wigner&Witmer,WWEnglish} diatomic eigenfunction.


\subsection{Air Wavelength vs. Vacuum Wavenumber}

For NMT analysis the recorded, digital intensity values versus calibrated wavelength are utilized. The variation of the refractive index, $r_i$, of air at 15~$^\circ$C, 101,325 Pa, and 0\% humidity, with wavenumber \cite{Ciddor},

\begin{equation}
10^8(r_i -1) = \frac{k_1}{(k_0 - \upsigma^2)} + \frac{k_3}{(k_2- \upsigma^2)},
\label{equation2}
\end{equation}
where $\upsigma$ is the wavenumber in units of $\upmu$m$^{-1}$, allows one to compute air wavelengths from the vacuum wavenumbers.  Table~\ref{table1new} lists constants in Eq.~(\ref{equation2}).

\begin{table}[h]
\caption{Constants for variation of refractive index, see Equation~\ref{equation2}.\label{table1new}}
\newcolumntype{C}{>{\centering\arraybackslash}X}
\begin{tabularx}{0.6\textwidth}{CC}
\hline
\textbf{Parameter}& \textbf{{Value ($\upmu$m$^{-2}$)}} 	 \\
\hline
{$k_0$ }& {238.0185} \\
{$k_1$ }& {5,792,105}  \\
{$k_2$ }& {57.362}   \\
{$k_3$ }& {167,917}   \\
\hline
\end{tabularx}
\end{table}

\section{Results}

This section elaborates analysis of recorded AlO spectra of the ${\rm B}\  ^2\,\Sigma^+ \longrightarrow {\rm X} \ ^2\,\Sigma^+$, $\Delta {\rm v}\,=\,0,\, \pm 1, \pm 2, 3$ sequences and progressions.
The use of ExoMol data and computed sets of line strength data that appear to be in use for extragalactic studies \cite{ExoMol} would alleviate computation of specific transitions that are investigated in laser-plasma laboratory experiments. The ExoMol database shows 4,945,580 transitions and 94,862 states including the three lowest electronic states, ${\rm X}~^2\Sigma^+, {\rm A}~^2\Pi, {\rm B}~^2\Sigma^+, {\rm C}~^2\Pi, {\rm D}~^2\Sigma^+,{\rm and} {\rm E}~^2\Delta$, e.g., 54,585 A states and 10,781 B states. The 10,781 B states lead to 774,575 B-X transitions.

The AlO-lsf B-X data contain 33,484 transitions. The differences in number of transitions are in part due to the number of rotational states, the cutoffs for Einstein A-coefficients and associated line strengths (see Eq.~(\ref{equationA})), or the establishment of sets of computed molecular parameters that fit data from high-resolution, Fourier-transform spectroscopy. The line positions are determined from high-resolution data with a standard deviation comparable to the estimated experimental errors of the high resolution line positions. The obtained, simulated line position accuracies are typically better than 0.05 ${\rm cm}^{-1}$.

{ In this work, a Gaussian profile models the spectrometer and intensified linear-array detector transfer function. However, a measured system transfer function or a Voigt function can replace the selected Gaussian profile provided that changes are implemented in the MATLAB source scripts for the recently communicated BESP and NMT scripts \cite{Foundations1}. N.B., the PGOPHER program allows one to accomplish Voigt profile fits.

{
\subsection{Analysis with NMT Program and ExoMol Line Strengths}
}
The AlO B-X line positions and Einstein A-coefficients (that are converted to line strengths) are collected in a data file that is compatible with the mentioned NMT-spectral fitting program. Figure \ref{figure1} illustrates spectra determined from temperature fitting with constant Gaussian line-width, $\Delta \uplambda$. The simulated spectrum utilizes only AlO B-X transitions in the experimental range of 430 nm to 540 nm. Analysis of the measured data with the AlO-lsf data \cite{Dors} reveals a temperature of ${\rm T} = 3,329\, {\rm K}$, and a fitted FWHM of 1\,nm (43\,cm$^{-1}$).

\begin{figure}[h!]
\centering
\includegraphics[width=0.975\textwidth]{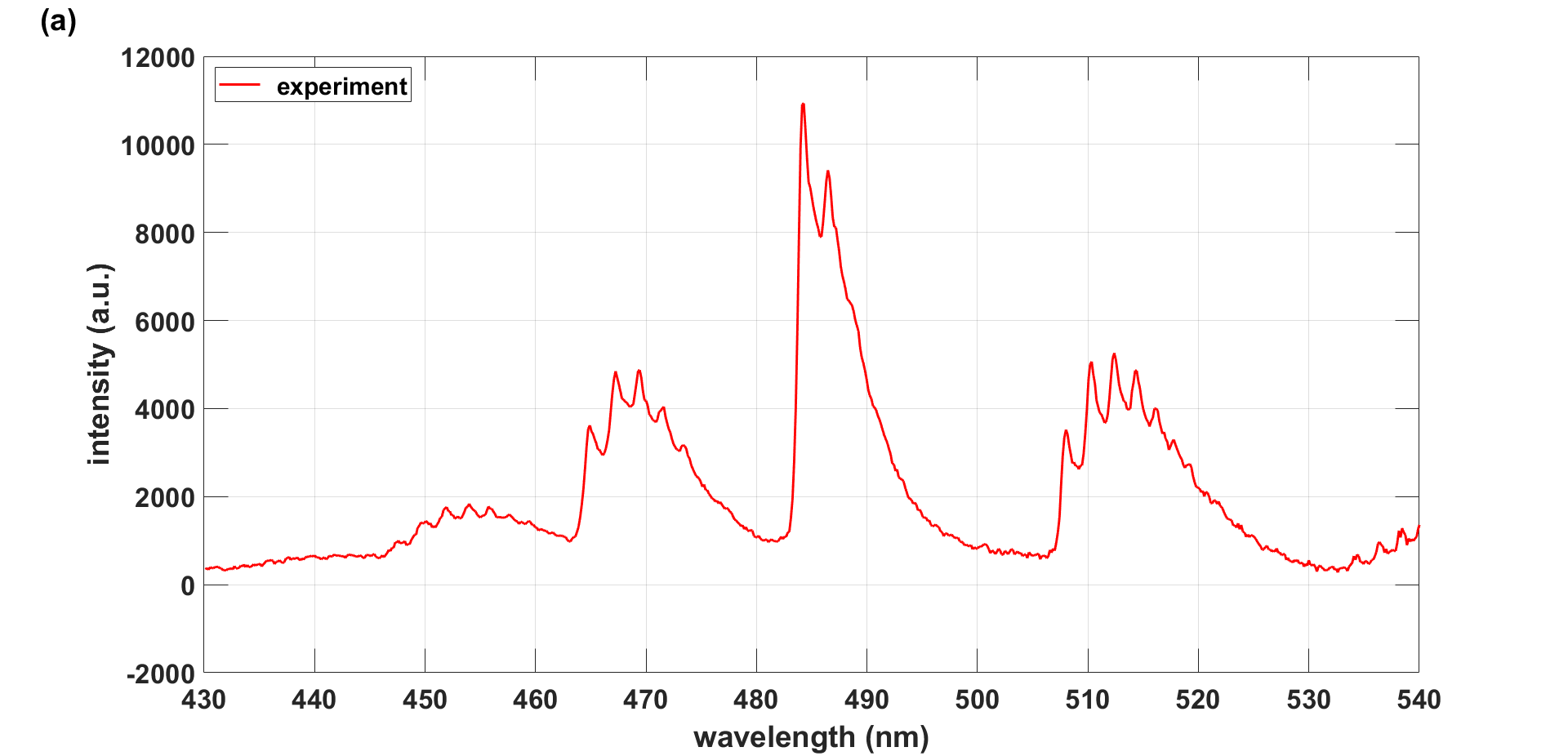} \includegraphics[width=0.975\textwidth]{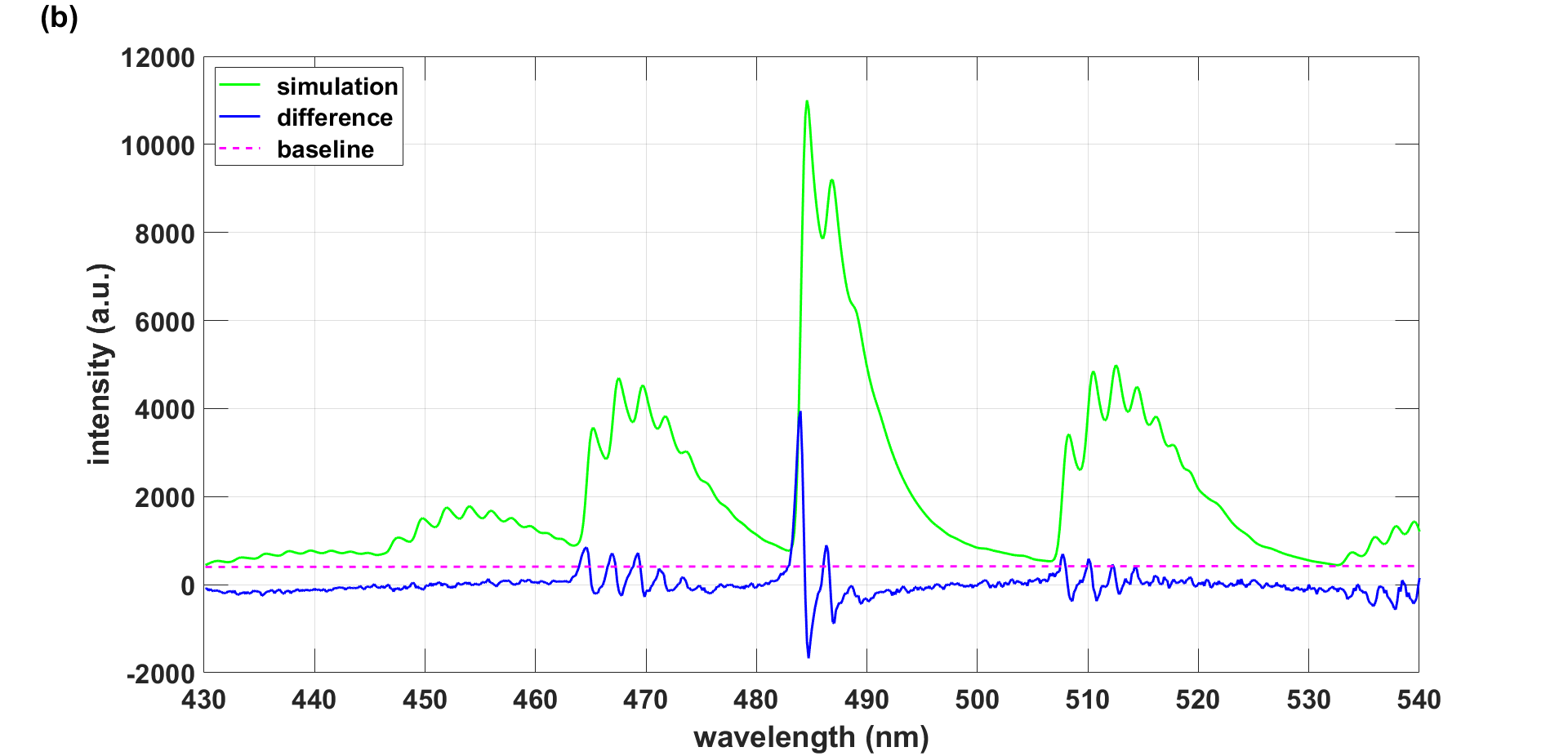}
\caption{{\textbf{(a)}} Experiment. {\textbf{(b)}} NMT fitting with ExoMol B-X data, ${\rm T} = 3,380\, {\rm K}$, $\Delta \uplambda = 1.0\ {\rm nm}$.}
\label{figure1}
\end{figure}

{
\subsection{Exomol AlO and  AlO-lsf Data Comparisons}
}

The simulated spectra are composed of quite a few individual rotational-vibrational transitions of the AlO B-X $\Delta {\rm v} = 0, \pm 1, \pm 2, + 3$ sequences and progressions. Tables \ref{tablelines} and \ref{tablelines2} summarize the number of lines in the data files.


\begin{table}[h]
\caption{Number of B-X transitions and those in the experimental range 430 nm to 540 nm (18,500 ${\rm cm}^{-1}$ to 23,250 ${\rm cm}^{-1}$). }\label{tablelines} 
\newcolumntype{C}{>{\centering\arraybackslash}X}
\begin{tabularx}{0.75\textwidth}{CCC}
\hline
\textbf{Database} & \textbf{AlO B-X} &\textbf{AlO B-X  430 nm to 540 nm}	 \\
\hline
ExoMol  & 774,575 & 169,143   \\
AlO-lsf & 33,484  & 29,258 \\
\hline
\end{tabularx}
\end{table}

\begin{table}[h]
\caption{Number of transitions in the experiment range 430\,nm to 540\,nm (see Tab. \ref{tablelines}) with Einstein A-coefficients, A$_{\rm coeff}$, larger than 10$^3$~s$^{-1}$.}\label{tablelines2}
\newcolumntype{C}{>{\centering\arraybackslash}X}
\begin{tabularx}{0.8\textwidth}{CC}
\hline
\textbf{Database} & \textbf{AlO B-X  430 nm to 540 nm  A$_{\rm coeff} > 10^3 s^{-1}$ }	 \\
\hline
ExoMol  & 104,260     \\
AlO-lsf & 29,258      \\
\hline
\end{tabularx}
\end{table}

\noindent Table \ref{tablecomparisons} displays agreements of lines within the indicated wavenumber range and otherwise the same identification for upper and lower levels of the transitions.

\begin{table}[h]
\caption{Subset AlO B-X lines of the ExoMol data that agree within $\Delta {\tilde \upnu}$ of 29,258 AlO B-X transitions in the AlO-lsf data for the experiment range 430\,nm to 540\,nm (or exactly 18,500~${\rm cm}^{-1}$ to 23,250 ${\rm cm}^{-1}$).} \label{tablecomparisons} 
\newcolumntype{C}{>{\centering\arraybackslash}X}
\begin{tabularx}{0.95\textwidth}{CCCCCCC}
\hline
\textbf{Database} & \textbf{$\Delta {\tilde \upnu} < 0.05\, {\rm cm}^{-1}$ }& \textbf{$\Delta {\tilde \upnu} < 0.2\, {\rm cm}^{-1}$}  & \textbf{ $\Delta {\tilde \upnu} < 1.0\, {\rm cm}^{-1}$} & \textbf{$\Delta {\tilde \upnu} < 2.0\, {\rm cm}^{-1}$}	&\textbf{$\Delta {\tilde \upnu} < 10.0\, {\rm cm}^{-1}$} &\textbf{$\Delta {\tilde \upnu} < 20.0\, {\rm cm}^{-1}$} \\
\hline
ExoMol  & 747   &   3,146 & 10,843 & 14,425 & 21,036 & 22,609 \\
\hline
\end{tabularx}
\end{table}

The differences in accuracy of the line positions can cause systematic errors in analysis of plasma emission spectra. Visualization of these differences is suggested by (a) generating a ``numerical experiment" spectrum using the AlO B-X data as extracted from the Exomol database, and then (b) analyzing the synthetic spectrum with the AlO-lsf database. Figure\,\ref{figure2} exhibits the Exomol-database generated and AlO-lsf line strength data analyzed results.
\begin{figure}[h!]
\centering
\includegraphics[width=0.975\textwidth]{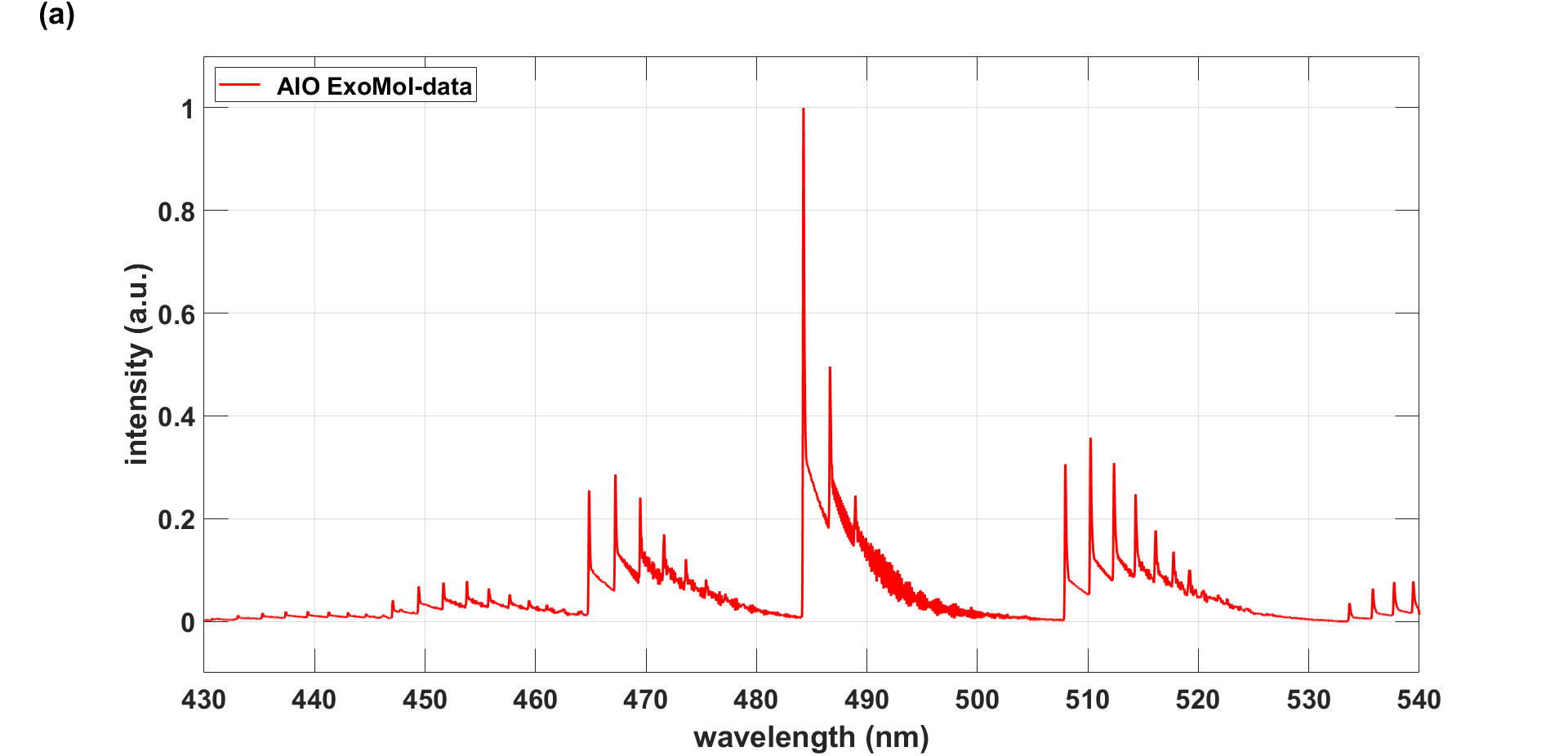} \includegraphics[width=0.975\textwidth]{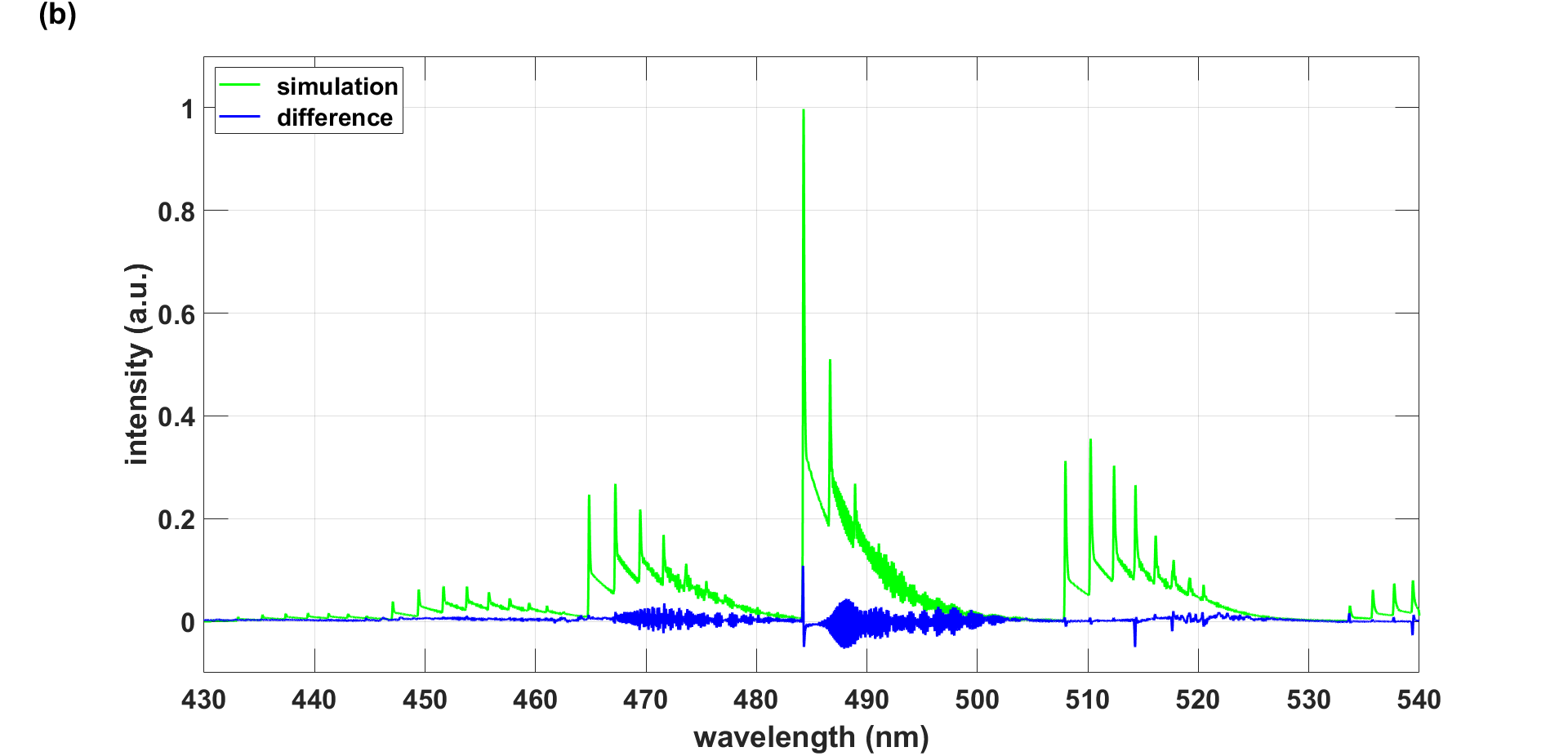}
\caption{{\textbf{(a)}} Numerical experiment data, T = 3,380\, K, $\Delta {\bar \uplambda} = 0.1\ {\rm nm}$. {\textbf{(b)}} NMT fitting with AlO-lsf B-X data, inferred temperature from fixed line-width fitting: ${\rm T} = 3,200\, {\rm K}$.}
\label{figure2}
\end{figure}

\noindent The obvious undulations in the difference spectrum illustrates the ExoMol inaccuracies indicated in Table\,\ref{tablecomparisons}. A temperature of T = 3,380 K and a full-width-at-half-maximum, fixed Gaussian line-width, $\Delta {\bar \uplambda}$, of 0.1 nm is selected for the ``numerical experiment.'' Analysis by only fitting temperature yields T = 3,200 K, i.e., a temperature that is about five per cent lower than specified for the spectrum in Fig.\,\ref{figure2}\,(a).

Further comparisons of the AlO-lsf and AlO-ExoMol databases explore the $\Delta {\rm v} = 0$ AlO B-X sequence. Figure~\ref{figure3} illustrates AlO-Exomol data computed for a temperature of 3,380\,K and a spectral resolution of 0.07\,nm, and it also shows the NMT-simulated results when only fitting temperature. As expected, there is a difference of approx. 30 per cent between specified (3,380\,K) and fitted temperature (2,460\,K). For a spectral resolution of 0.1\,nm, the fitted temperature for the $\Delta {\rm v} = 0$ AlO B-X sequence equals 2,920\,K, or in other words, the temperature difference decreases is approx. 14 per cent lower than specified.

Tables~\ref{table5} and \ref{table6} summarize comparisons of the transition lines with Einstein A-coefficients larger than 10$^3$~s$^{-1}$. There are about five times more lines in the ExoMol database for the 10-nm spectral window. Among the 2,818 AlO-lsf lines only 96 ExoMol lines agree within better than $0.05\, {\rm cm}^{-1}$, and 1,517 ExoMol lines show wave numbers within $3\, {\rm cm}^{-1}$ (about 0.07\,nm) of those of the AlO-lsf data.

\begin{table}[h]
\caption{Number of transitions in the experiment range 483\,nm to 493\,nm with Einstein A-coefficients, A$_{\rm coeff}$, larger than 10$^3$~s$^{-1}$.}\label{table5}
\newcolumntype{C}{>{\centering\arraybackslash}X}
\begin{tabularx}{0.8\textwidth}{CC}
\hline
\textbf{Database} & \textbf{AlO B-X 483 nm to 493 nm A$_{\rm coeff} > 10^3 s^{-1}$ }	 \\
\hline
ExoMol  & 10,159     \\
AlO-lsf & 2,818      \\
\hline
\end{tabularx}
\end{table}

\begin{table}[h]
\caption{AlO B-X lines of the ExoMol data that agree within $\Delta {\tilde \upnu}$ of 2,818 transitions in the AlO-lsf data for the range 483\,nm to 493\,nm (20,284~${\rm cm}^{-1}$ to 20,704 ${\rm cm}^{-1}$).} \label{table6} 
\newcolumntype{C}{>{\centering\arraybackslash}X}
\begin{tabularx}{0.75\textwidth}{CCCC}
\hline
\textbf{Database} & \textbf{$\Delta {\tilde \upnu} < 0.05\, {\rm cm}^{-1}$ }& \textbf{$\Delta {\tilde \upnu} < 0.3\, {\rm cm}^{-1}$}  & \textbf{ $\Delta {\tilde \upnu} < 3.0\, {\rm cm}^{-1}$} \\
\hline
ExoMol  & 96   &   506 & 1,517 \\
\hline
\end{tabularx}
\end{table}

The AlO-lsf line strength database has been extensively tested \cite{Pariggerbook}. The ExoMol database appears acceptable within $\simeq$ 20 wavenumbers, i.e., average spectral resolution of $\simeq$ 0.3~nm. Analysis of higher than 0.3\,nm resolution data, viz. spectral resolution of 0.07\,nm, is affected by the inaccuracies of the line positions listed in the ExoMol database.
The AlO-lsf database accuracy is better than 0.05\,cm$^{-1}$ that corresponds to $\simeq$\, 1\,picometer for the AlO B-X bands.

\begin{figure}[h!]
\centering
\includegraphics[width=0.975\textwidth]{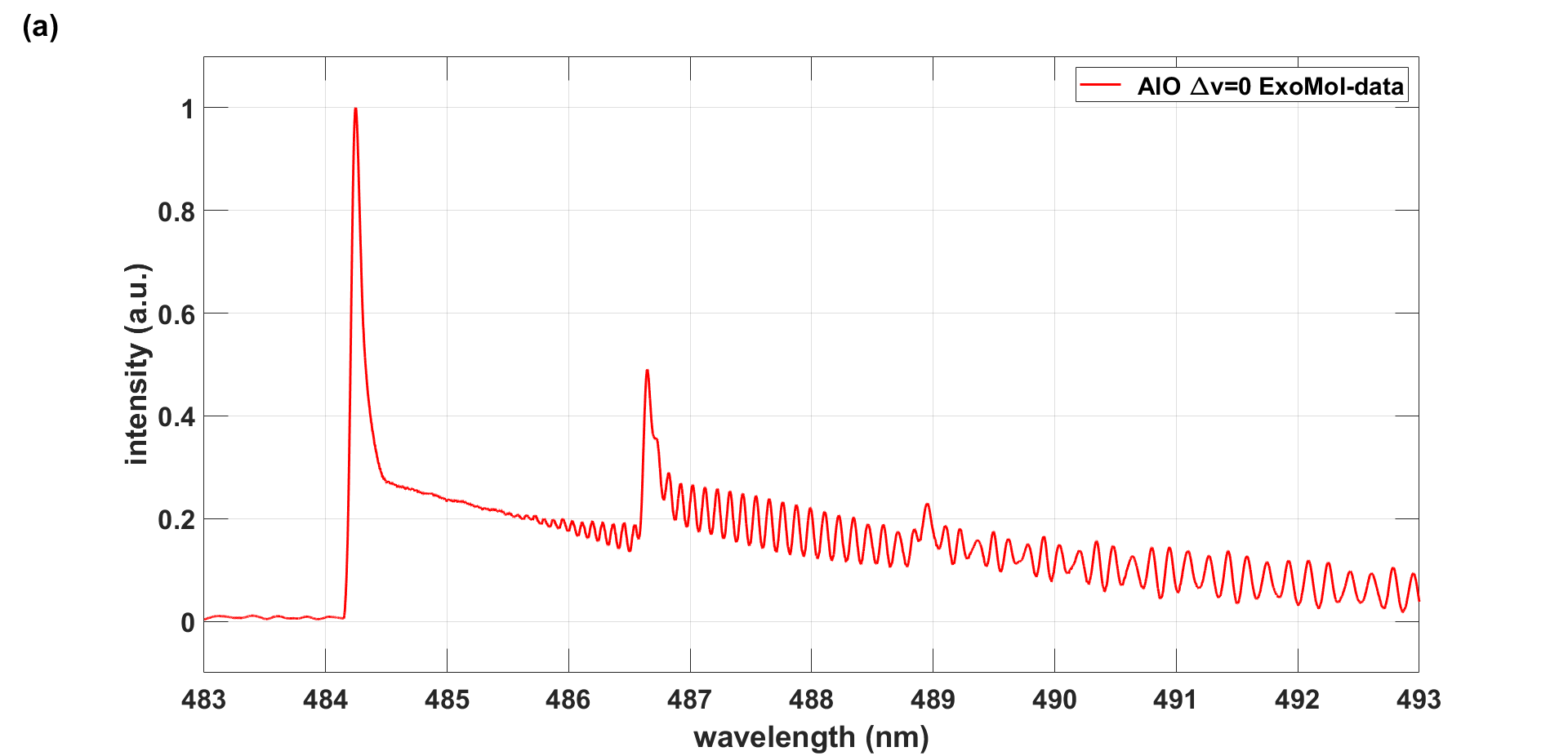} \includegraphics[width=0.975\textwidth]{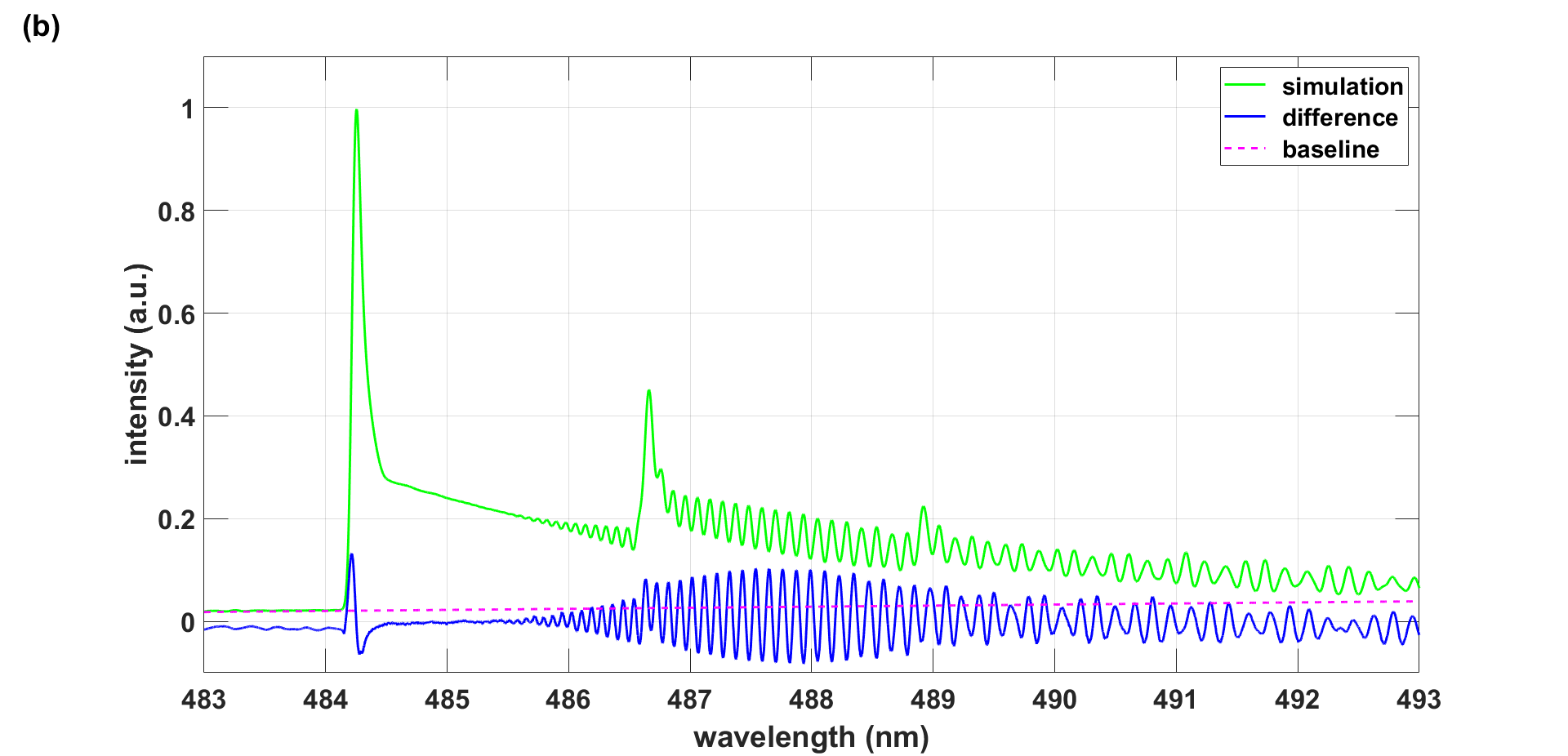}
\caption{{\textbf{(a)}} Numerical experiment data, T = 3,380\, K, $\Delta {\bar \uplambda} = 0.07\ {\rm nm}$. {\textbf{(b)}} NMT fitting with AlO-lsf B-X data, inferred temperature from fixed line-width fitting: ${\rm T} = 2,460\, {\rm K}$.}
\label{figure3}
\end{figure}

\section{Discussion}

The AlO $B \, ^2\Sigma^+ - X \, ^2\Sigma^+$, $\Delta {\rm v} = 0, \pm 1, \pm 2, + 3$ sequences and progressions reveal many vibrational and rotational transitions that are usually not individually resolved in the study of laser-induced plasma emissions in the spectral range of 430 nm to 540 nm. Analysis of the 1-nm spectral resolution experimental emission spectrum with ExoMol line strengths and the NMT program shows AlO excitation temperature of $\simeq$ 3,380 K that is consistent with previous analysis with AlO-lsf line strengths. However, for the $\Delta {\rm v} = 0$ sequence, and for a spectral resolution of 0.07\,nm, there is a significant difference of ExoMol-computed and AlO-lsf fitted temperatures of typically 30\%.

The agreement of the ExoMol AlO B-X and AlO-lsf line position is marginal when using accuracies of the order of 0.05\,cm$^{-1}$, or of the order of 1 picometer. For spectral resolutions of 3\,cm$^{-1}$, or about 0.07\,nm, and  in the $\Delta {\rm v} = 0$ sequence, use of the AlO-lsf database is recommended. For measurements with spectral resolutions of 43\,cm$^{-1}$, or of the order of 1 nanometer, almost exactly identical results are inferred from fitting of measured ablation spectra. A significant advantage of the AlO-lsf database is its accuracy in predicting line position compared to the ExoMol database. The AlO-lsf line strength table is generated by fitting high resolution Fourier-transform data rather than computation from first principles.




\acknowledgments{\noindent The author (CGP) acknowledges the support in part by the Center for Laser Applications at the University of Tennessee Space Institute.}

%

\section*{References}


\end{document}